\documentclass[%
reprint,
amsmath,amssymb,
aps, groupedaddress,superscriptaddress
]{revtex4-2}
\usepackage[usenames,dvipsnames]{xcolor}
\usepackage{amssymb}
\usepackage{graphicx}
\usepackage{amsmath}
\usepackage{comment}
\usepackage{upgreek}
\usepackage{bbm}
\usepackage[bookmarks=true,colorlinks,citecolor=blue,urlcolor=blue]{hyperref}
\usepackage{braket}
\usepackage{babel}
\usepackage{blindtext}
\usepackage[normalem]{ulem}
\usepackage{physics}

\usepackage{mathtools}

\definecolor{mypurp}{rgb}{0.35, 0, 0.7}

\usepackage{amsthm}
\usepackage{txfonts}

\begin{document}

%\title{Passive error correction in protected superconducting qubits via analogies to cat qubits}
%\title{Viewing fluxonium through the lens of the cat qubit}
\title{Viewing protected superconducting qubits through the lens of the cat qubit}

\def\aws{AWS Center for Quantum Computing,  Pasadena, California 91125, USA}

\author{Simon Lieu}
\email{slieu@amazon.com}
\author{Emma L.~Rosenfeld}
\thanks{Current affiliation: Google Research}
\author{Kyungjoo Noh}
\author{Connor T.~Hann}

%\thanks{ ...}

\affiliation{\aws}

\date{\today}
\begin{abstract}

    We draw analogies between  protected superconducting  qubits  and bosonic qubits by studying the fluxonium Hamiltonian in its Fock basis. The mean-field phase diagram  of fluxonium (at the sweet spot) is identified, with a region in parameter space that is characterized by $\mathbb{Z}_2$-symmetry-broken ground states.
    In the heavy fluxonium limit, these  ground states are well approximated by squeezed coherent states in a Fock basis (corresponding to persistent current states with  definite flux but indefinite charge),
    and  simple expressions are provided for them in terms of the circuit parameters. 
    We study the noise bias in fluxonium via a universal Lindblad master equation and find that the bit-flip rate is exponentially small in $E_j/(k_B T)$, while the phase-flip rate does not get worse with this ratio. 
    %This scaling is favorable compared to the cat qubit, which has a phase-flip rate that gets linearly worse with the number of photons.
    Analogous behavior is found in $\cos(2 \theta)$ qubits. We describe cat-qubit-inspired bias-preserving $X$ and $CX$ gates for fluxonium. We discuss first steps towards generating an Ising interaction between protected superconducting  qubits on a two-dimensional lattice,  with the aim of achieving a   passive quantum memory by coupling a static Hamiltonian to a generic thermal bath.

\end{abstract}

\maketitle

\section{Introduction}

Quantum error correction relies on redundantly encoding quantum information into a large Hilbert space such that physical errors do not directly cause logical errors \cite{lidar_book, kitaev2003, gottes1997, eric2002}. One popular approach is called ``hardware efficient bosonic encoding''  where information is   encoded into the infinite-dimensional Hilbert space of a harmonic oscillator \cite{Joshi_2021}, including qubits such as the GKP code \cite{gotte2001},  binomial code \cite{binomial2016}, and  cat code \cite{mirrahimi2014}.
For certain codes, such encodings are advantageous because there is  notion of distance in ``phase space'': codewords are well separated in phase space while errors  only perturb locally. Thus a certain level of protection can be achieved at the level of the physical hardware before applying ideas from active error correction. Notably, certain bosonic qubits (e.g.~the cat code) can \textit{exponentially} suppress the   bit-flip error rate by only linearly increasing separation in phase space. The cost is that such schemes  typically require 
%working with a harmonic oscillator (e.g.~photonic resonator or LC circuit) in the presence of 
external drives and/or engineered dissipation to confine a harmonic oscillator 
%(e.g.~photonic resonator or LC circuit) 
into a bosonic code space, which  
%. From an experimental point of view, this leads 
can lead to complexities due to drive-induced resonances \cite{putterman2024b, carde2024} 
%can lead to a significant hardware overhead 
and  nonequilibrium dynamics that can be difficult to model.
%and introduces additional moving parts compared to a ``trivial'' encoding into the lowest two Fock levels. 
%From a theoretical point of view, it is challenging to model such systems because of their driven-dissipative dynamics.

On the other hand, it is known that certain superconducting ``protected qubits''  can also exhibit hardware-level protection against noise, e.g.~an exponentially-long $T_1$ time (i.e.~bit-flip time) as a function of circuit parameters \cite{schuster2021}. Notably, this protection is achieved  \textit{without} an external drive or engineered dissipation. Canonical examples include: the fluxonium qubit \cite{manu_2009, zhang_2021, lin_2018, dogan_2023, alibaba, nguyen_2022, oliver_2023, hassani_2023}, zero-pi qubit \cite{kitaev:2006zeropi, groszkowski_2018,brooks_2013, ioffe:2002, houck2021, Paolo_2019, xanda_2025}, and $\cos(2\theta)$ qubit \cite{cos2t, larsen2020, vool2024, patel2024, pop_2008, bell_2014, dodge_2023}. The standard explanation for $T_1$ protection is  that most errors are caused due to environmental noise that couples to the charge or flux operator of the circuit, and the relevant transition matrix elements between logical codewords (i.e.~$\langle \bar{0 }| n |\bar{1 } \rangle$ and $\langle \bar{0 }| \phi |\bar{1 } \rangle$) are exponentially suppressed in a  ratio of energy scales (e.g.~the Josephson energy over capacitive energy: $E_j /E_c$ in fluxonium). 
%These circuits have the advantage that they can exponentially-suppress errors at the level of hardware

In this work we would like to  explain the protection mechanism of a superconducting qubit using the language of  bosonic encodings: The infinite-dimensional Hilbert space of a circuit hosts  ground states that are nearly degenerate and far away from each other in phase space. For concreteness, we will describe fluxonium in a language that draws an analogy with the cat code, emphasizing similarities and differences.

\begin{figure}
    \centering
    \includegraphics[scale=0.5]{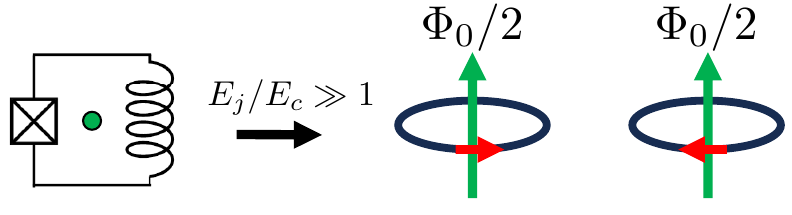}
    \caption{ The  limit of a robust  classical bit for fluxonium (infinitesimally-thin insulating barrier): A superconducting ring (black) with half of a flux quantum threaded through it (green) will  develop a persistent current (red) that \textit{either} flows in the clockwise or counterclockwise direction since the total flux through a superconducting ring must be quantized in units of $\Phi_0$. This is an example of $\mathbb{Z}_2$ spontaneous symmetry breaking that  protects a classical bit in a ``zero-dimensional'' system. Note that fluxonium does not require an external drive or engineered dissipation, and has finite energy even in this classical limit (in contrast to the cat code).}
    %\caption{ The  limit of a robust  classical bit for (a) the photonic cat code, (b) fluxonium. (a): A photonic cavity that is driven with a two-photon drive (and experiences two-photon loss) admits two stable coherent-state solutions which oscillate with the drive frequency and are $\pi$ out of phase with respect to each other. (The electric field at a given time for the two solutions is sketched in blue.) (b) A superconducting ring (black) with half of a flux quantum threaded through it (green) will develop a current (red) that \textit{either} flows in the clockwise or counterclockwise direction since flux must be quantized in units of $\Phi_0$ through a superconducting ring. Both models are examples of $\mathbb{Z}_2$ spontaneous symmetry breaking in a ``zero-dimensional'' system leading to a good classical bit. Note that fluxonium does not require an external drive (or engineered dissipation), and has finite energy.}
    \label{fig:ssb}
\end{figure}

The cat code is the canonical example of a bosonic code that features passive bit-flip noise suppression:  Information is encoded into two macroscopic coherent states: $|\pm \alpha \rangle$, which can be stabilized, e.g.,
%typically 
via a two-photon drive ($\lambda$) and engineered two-photon loss ($\kappa_2$) \cite{mirrahimi2014}.  In the limit of large $|\alpha|^2 \sim  \lambda/\kappa_2$, the codewords $|\pm\alpha\rangle$  are well separated in phase space, while the typical error mechanisms of single-photon loss and dephasing ($L \sim a, a^\dagger a$) only perturb states locally \cite{prx2019}.  Physically:  A cavity that is strongly driven with pairs of photons (and loses pairs of photons) will  stabilize to one of two classical states of light that are  $\pi$ out of phase; the classical coherent states  break photon parity: $\exp[i \pi a^\dagger a]|\alpha\rangle = |- \alpha\rangle$. A robust classical bit is associated with a phase in parameter space  characterized  by  symmetry-broken steady states \cite{lieu2020}.
%Thus one can view the spontaneous symmetry breaking of photon parity as responsible for this stable classical bit \cite{lieu2020}. 
%[See Fig.~\ref{fig:ssb}(a).]

In this work we show that fluxonium  has nearly-degenerate ground states that spontaneously break flux parity, corresponding to  \textit{squeezed} states  in the infinite-dimensional Hilbert space of its LC subsystem that have nearly-definite flux.
%. These correspond  to states with definite flux but indefinite charge that are well separated in phase space 
Thermal processes    will thus  evolve the system towards a nontrivial  bosonic state at low temperature. The separation of the two  codewords (ground states) in phase space will be mostly controlled by the ratio of two circuit parameters: $|\alpha '|^2  \sim \sqrt{E_j / E_c}$.
%, where $E_j$ is the Josephson energy and $E_c$ is the capacitive energy. 
%These squeezed states are the ground states of the model \textit{in the lab frame} (unlike the cat code).   
Physically these ``classical''  states in fluxonium correspond to persistent  current states that circulate clockwise or counterclockwise in the circuit. Scaling $|\alpha '|^2  \rightarrow \infty$ 
%thus does not  require  driving the system infinitely hard, but rather 
represents the limit where the physical Josephson barrier width (the insulator between two superconductors) goes to zero ($E_j \rightarrow \infty$). In the extreme limit of no physical barrier, the model represents a  superconducting ring  with half of a flux quantum ($\Phi_0/2$) threaded through it. Since the total flux through the loop must be quantized in units of $\Phi_0$, the superconducting ring forms a current in the clockwise or counterclockwise direction to induce an additional flux of $\pm \Phi_0/2$. [See Fig.~\ref{fig:ssb}.] Currents in a superconducting ring have been observed to persist without decay on a timescale on the order of years  \cite{crowe_1957, broom_1961, tsuei2000, gough1992}, suggesting that the upper bound on the classical-bit lifetime could be very long in practice.
%This behavior has been confirmed experimentally \cite{crowe_1957,tsuei2000, gough1992} and suggests that there is no upper limit to the lifetime of the classical bit. 
The current in this classical limit is \textit{finite}, only costing a finite energy (unlike  the cat code which requires an increasing number of photons, i.e.~large electric field). 
%(unlike  the cat code which requires an increasing electric field). 
Fluxonium thus provides a  limit where a single bosonic mode can reliably store a classical bit at finite energy.

%This produces a superconducting current state which  is equally stable when moving in the clockwise or counterclockwise directions  since the total flux through the loop must be quantized in units of the flux quantum ($\Phi_0$)

Beyond protected superconducting qubits, the field of passive error correction more generally aims to identify ways to suppress errors at the level of physical hardware 
%Passive error correction comes in many flavors 
\cite{bacon2006, yoshida2011, roberts2020b, terhal2015, brown2016, bombin2013, haah2011, bravyi2013, loss2010, breuckmann2016,  Alicki2010, passive3, passive4, passive5, passive7,  pastawski2011, sala_2024,wang_2025, kapit2016, reiter2017,  brell2014, ding2024, young2024, ackermann2024, morales2024b,  hseih2024, zhu2024, tibor2024, hong2024, rojkov2024, le_2019, gravina_2023, jerry_2025, lake2024}.  Exponential suppression of bit and/or phase-flip errors without measurements is typically associated with nontrivial phases \cite{eric2002}, necessary for robust steady-state degeneracy in the presence of noise \cite{liu2023dissipative}. Notable examples include  a protected \textit{classical bit} due to thermally-stable $\mathbb{Z}_2$ spontaneous symmetry breaking in the 2D Ising model, and a protected \textit{qubit} in the 4D toric code due to thermally-stable topological order. 
%Our work draws a connection  between the fields of bosonic error correction and superconducting protected qubits by showing that both can host a protected classical bit due to $\mathbb{Z}_2$ spontaneous symmetry breaking   in an infinite-dimensional Hilbert space.

A recent proposal \cite{lieu2023candidate} suggests that a passive quantum memory can be achieved in a 2D lattice of bosonic cat qubits coupled via nearest-neighbor Ising interactions,
%by constructing a 2D Ising model out of bosonic cat qubits on each lattice site, 
i.e.~a repetition cat code where the repetition code is done passively. In this work, we propose an analogous scheme to achieve a passive qubit via a 2D Ising model built out of $\cos(2 \theta)$  qubits. The main advantage of the proposal outlined here vs the one in Ref.~\cite{lieu2023candidate} is twofold: (1)  It only relies on generic thermal coupling for protection, rather than a driven-dissipative (nonequilibrium) environment; (2) There is finite energy on each lattice site even in the  classical-bit limit, thus eliminating any   anomalous behavior associated with large photon numbers (e.g.~drive-induced heating). The key ingredient needed to bypass  self-correcting no-go theorems \cite{bravyi2009, brown2016} is an infinite-dimensional Hilbert space on each lattice site, which arises naturally for superconducting qubits. 

\section{Phase diagram  and ground state characterization}

\begin{figure}
    \centering
    \includegraphics[scale=0.35]{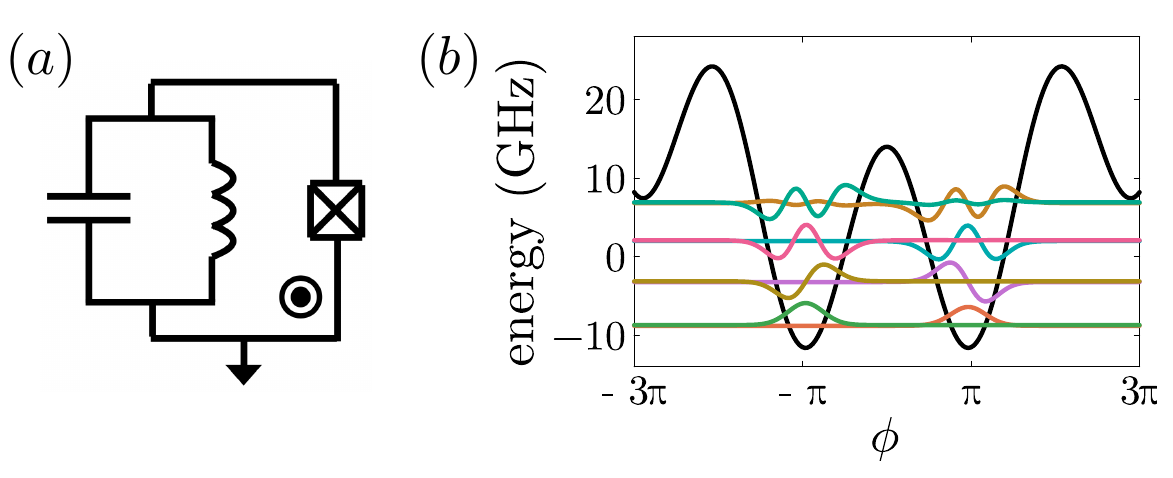}
    \caption{(a) Circuit diagram for fluxonium: a capacitor, inductor, and Josephson junction are connected in parallel, with an external flux $\phi_e$ through the loop. We will be working at or near the sweet spot $\phi_e = \pi$. (b) Flux potential and lowest-energy eigenstates for fluxonium in the heavy limit slightly off of the sweet spot: $E_j/ h = 14 \text{ GHz}, E_c / h = 0.3 \text{ GHz} , E_l / h = 0.5 \text{ GHz} $, $ \phi_e = 0.99 \pi $. }
    \label{fig:fluxonium}
\end{figure}

Consider the fluxonium Hamiltonian, i.e.~a capacitor, inductor, and Josephson junction connected in parallel, with an external flux through part of the loop. [See Fig.~\ref{fig:fluxonium}(a).] The Hamiltonian reads:
\begin{equation}\label{eq:fluxonium_ham}
    H = 4 E_c n^2 + \left( \frac{E_l}{2} \right) \phi^2 - E_j \cos \left(  \phi - \phi_{e} \right)
\end{equation}
where $\phi$ is the superconducting phase, $n$ is the (dimensionless) charge operator, $E_c, E_l, E_j$ are the capacitive, inductive, and Josephson energies of the circuit, and $\phi_e$ is proportional to the flux through the loop. The Hamiltonian can be rewritten in terms of creation and annihilation operators of excitations in the LC subsystem:
\begin{equation}\label{eq:fluxonium_ham_fock}
    H = \hbar \omega a^\dagger a + E_j \cos(  \phi_0 (a + a^\dagger) )
\end{equation}
where $\phi_0 = \left(  2 E_c /E_l\right)^{1/4}$, $ \hbar \omega = \sqrt{8 E_l E_c}$, and we have set $\phi_e = \pi$ (leading to the $+$ sign in front of the Josephson term) which corresponds to half of a flux quantum and is called the sweet spot since the Hamiltonian is first-order insensitive to flux noise. The Josephson term does not commute with the harmonic oscillator term, ensuring that  the ground state is not just the vacuum of photons.

The fluxonium Hamiltonian has a $\mathbb{Z}_2$ parity symmetry: $P = e^{i \pi a^\dagger a}, [H,P]=0$, since all terms are even in powers of $a,a^\dagger$. (This corresponds to inversion symmetry in the flux potential around $\phi=0$.) We can ask whether the ground state of the Hamiltonian respects this symmetry as a function of circuit parameters, i.e.~$P | \text{gnd} \rangle \propto |\text{gnd} \rangle$ in the symmetric case, and  $ P |\text{gnd}  \rangle \not\propto |\text{gnd}  \rangle$ in the symmetry-broken case.

We guess a symmetry-broken solution for the ground state, in the form of a squeezed coherent state:
\begin{equation}
 |\alpha , \theta \rangle = D(\alpha) S(\theta)|\text{vac}\rangle
\end{equation}
where
\begin{equation}
    D(\alpha) = \exp[ \alpha a^\dagger - \alpha^* a ], \quad  S(\theta) = \exp[  (\theta^* a^2 - \theta a^{\dagger 2} )  /2 ]
\end{equation}
is the displacement operator and  the squeezing operator respectively, and $|\text{vac}\rangle$ is the vacuum of photons.  This choice corresponds to a gaussian wavefunction in flux space, with a mean and variance that depends on the parameters $\alpha, \theta$ respectively. [See Fig.~\ref{fig:fluxonium}(b).] The associated mean-field energy is
\begin{align}
    E_{mf} &=  \langle \alpha , \theta |  H|\alpha , \theta \rangle  \\
    &= \hbar \omega(\alpha^2 + \sinh^2(\theta)) + E_j  \exp[-\phi_0^2 e^{-2\theta}/2] \cos(2 \alpha \phi_0)
\end{align}
Note that this expression is invariant under $\alpha \rightarrow - \alpha$ which means that   two solutions   minimize the energy  if the  optimal $\alpha$ has nonzero magnitude. (The symmetry-broken phase is characterized by such   solutions.)

\begin{figure}
    \centering
    \includegraphics[scale=0.4]{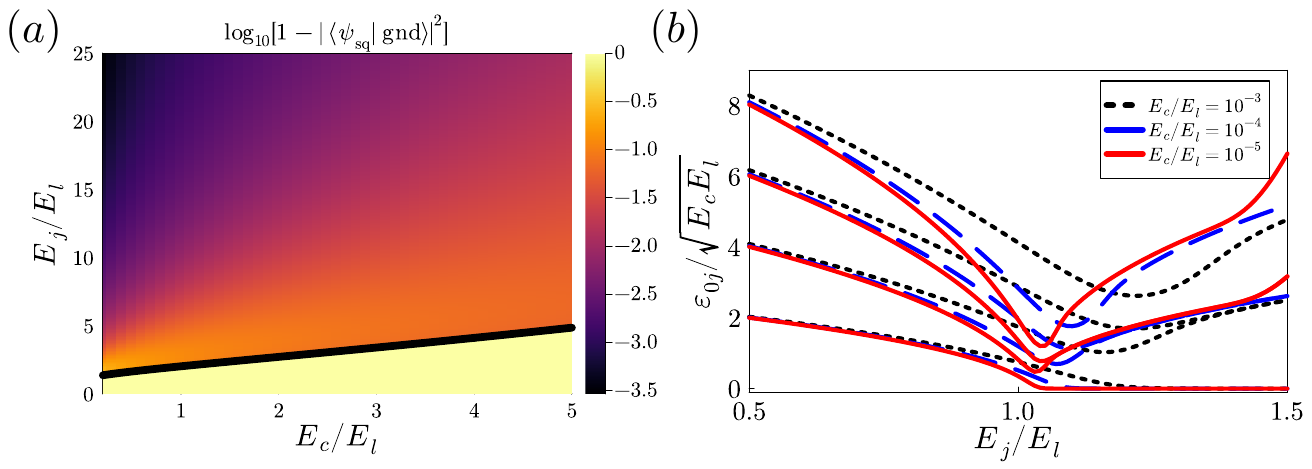}
    \caption{ (a) Black line: Boundary defined by the line in Eq.~\eqref{eq:boundary}; above the line, the symmetry-broken solutions start to become energetically favorable compared to the symmetric (trivial) state. Above the line: Overlap between the exact ground state of fluxonium at the sweet spot and a symmetric superposition of squeezed  states: $|\psi_{\text{sq}} \rangle \sim |\alpha, \theta \rangle + | - \alpha, \theta \rangle$   with $\alpha, \theta$ defined in Eqs.~\eqref{eq:alpha} and \eqref{eq:theta}.  The ansantz agrees well for  heavy fluxonium, $E_j / E_c \gg1$. (b) Lowest four energies above the ground state ($\epsilon_{0j}$) across the phase boundary: As the thermodynamic limit is approached ($E_c \rightarrow 0$), the spectrum closes at the boundary ($E_j=E_l$) and the first excited state is exponentially small  in the symmetry-broken phase.}
    \label{fig:overlap}
\end{figure}

We numerically optimize the expression for $E_{mf}$ over the parameters $\alpha, \theta$ and find  a second-order boundary where both $\alpha, \theta$
 grow smoothly from zero across the boundary. To find an analytic expression for this line we thus 
%We look for a second-order phase transition, defined by a line where the values of $\alpha, \theta$ start to acquire a nonzero expectation value across the boundary. 
%We can find the phase diagram of fluxonium by minimizing $E_{mf}$ over $\alpha, \theta$ for each choice of circuit parameters ($E_c, E_l, E_j$) and asking  whether we obtain a nonzero solution for $\alpha, \theta$.  
expand the  expression for $E_{mf}$ in powers of $\alpha$ (setting $\theta=0$):
\begin{equation}
    E_{mf} \approx E_j e^{-\phi_0^2/2}  + (\hbar \omega - 2 \phi_0^2 E_j e^{- \phi_0^2/2}) \alpha^2 +  \left( E_j e^{-\phi_0^2/2 } \frac{ (2 \phi_0)^4 }{4!} \right) \alpha^4 .
\end{equation}
 $\alpha$  acquires a nontrivial solution when the coefficient of $\alpha^2$ is negative, corresponding to the line:
\begin{equation}\label{eq:boundary}
        E_j / E_l = \exp[\sqrt{2 E_c / E_l}/2] .
\end{equation}
This is boundary is plotted in Fig.~\ref{fig:overlap} and agrees with numerical optimization of $E_{mf}$.

In the limit where $E_j >E_l$ and $E_c \rightarrow 0$ (i.e.~the limit of heavy fluxonium), we expect that the lowest mean-field energy is found by minimizing the energy of the Josephson term. To this end we guess  $\alpha = \pi / (2 \phi_0) - \delta$ and minimize over $\delta$, i.e.~setting $d E_{mf}/d \delta =0$. We find an optimal $\alpha$ at:
\begin{equation}
     \alpha = \frac{\pi }{2 \phi_0} \left(  1 - \frac{ \hbar \omega}{ \hbar \omega + E_j (2 \phi_0^2) e^{- \phi_0^2/2} }  \right).
\end{equation} 
In the limit of $E_c \rightarrow 0$ this expression reduces to:
\begin{equation}  \label{eq:alpha}
         \alpha \approx  \frac{\pi }{2} \left( \frac{E_l}{2 E_c} \right)^{1/4} \left( \frac{E_j}{ E_j + E_l } \right).
\end{equation}
%We find that  we can achieve arbitrary separation in phase space by tuning $E_c \rightarrow 0$ (which implies a large capacitance since $E_c \sim 1/C$ where $C$ is the capacitance). Note that these large coherent states are the \textit{ground states} of the model, i.e.~we can achieve large separation in phase space without pumping energy into the system, unlike the photonic cat code.

Similarly, we can find the optimal squeezing parameter by setting: $d E_{mf}/d \theta = 0$ and solving for $\theta$. If we specify the value for $\alpha$ in Eq.~\eqref{eq:alpha}, we find the following expression for the optimal squeezing parameter (which again should be good in the limit $E_c \rightarrow 0$):
\begin{equation}
    \theta = \frac{1}{4} \ln \left( \frac{E_j}{ E_l} -\frac{\pi^2 E_j E_l}{2(E_j + E_l)^2} +1 \right).
\end{equation}
In the limit $E_j \gg E_l, E_c$ this expression can be further simplified to
\begin{equation} \label{eq:theta}
    \theta \approx \frac{1}{4} \ln \left( \frac{E_j}{ E_l} \right).
\end{equation}

The expressions for $\alpha, \theta$ provide a simple characterization for the ground states of fluxonium near the sweet spot as a function of the circuit parameters.  We note that any finite parity-preserving Hamiltonian has eigenstates with definite parity and hence to benchmark the accuracy of our analysis we compare the exact ground state of the Hamiltonian \eqref{eq:fluxonium_ham_fock} at the sweet spot with the symmetric state: $|\psi_{\text{sq}} \rangle \sim |\alpha, \theta \rangle + | - \alpha, \theta \rangle$, with $\alpha, \theta$ given in Eqs.~\eqref{eq:alpha} and \eqref{eq:theta} respectively.  This is plotted in  Fig.~\ref{fig:overlap}(a); we find excellent agreement for a wide region of parameter space.  The agreement gets exponentially better in the heavy fluxonium limit where $E_j/E_c \gg 1$.
%$E_j/ E_l \gg 1$ or $E_c/ E_l \ll 1$.

\begin{figure}
    \centering
    \includegraphics[scale=0.2]{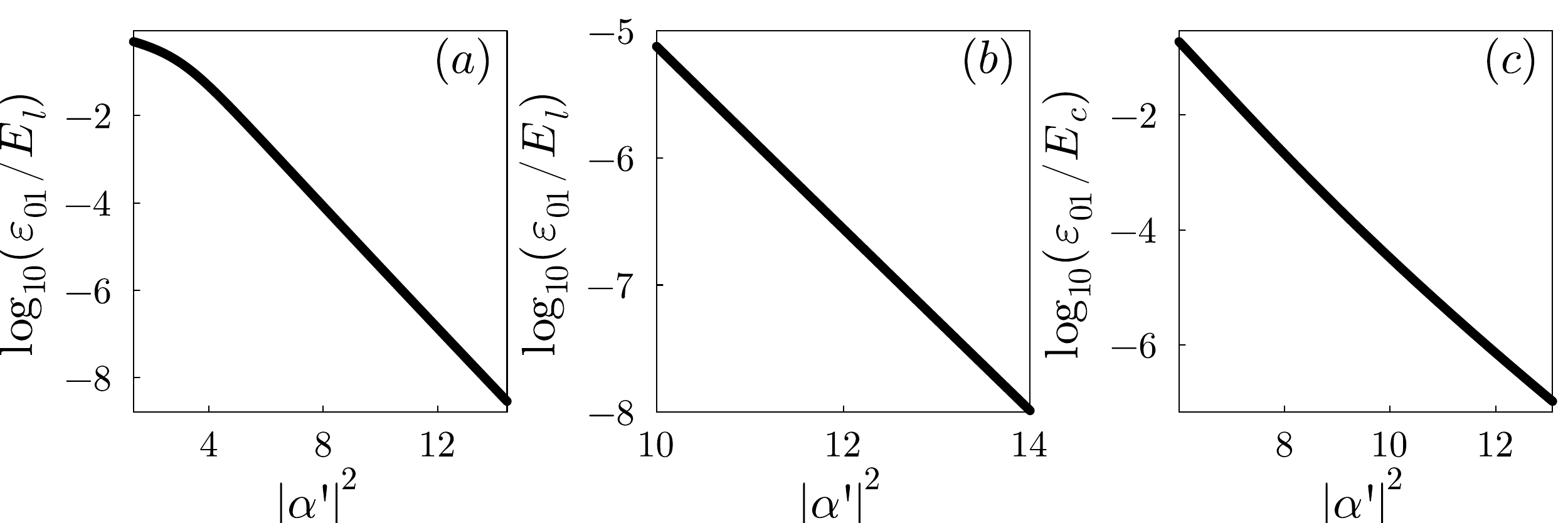} 
    \caption{ Exponentially-small splitting between the lowest two eigenstates of fluxonium $\epsilon_{01}$ (on a log plot) at the sweet spot as a function of the thermodynamic variable $N = |\alpha'|^2$, see Eq.~\eqref{eq:alpha_p}. (a) $E_c /E_l = 0.1, E_j/E_c \in [1,10]$, (b) $ E_j/E_l = 10, E_c/E_l \in [0.1,3]$, (c) $E_j/E_c= 60, E_l/E_c \in [10,30]$.}
    \label{fig:splitting}
\end{figure}

Formally a phase diagram should be defined in the thermodynamic limit.
%For fluxonium, $N\rightarrow\infty$ as $E_c\rightarrow0$ which suggests that the vertical  line characterized by $E_c/E_l=0$ in Fig.~\ref{fig:overlap} is the true phase diagram; by tuning $E_c$ one can tune towards the thermodynamic limit, analogous to adding more lattice sites in the transverse-field Ising model.
%While finite systems have symmetric eigenstates, symmetry-broken phases are characterized 
%any finite symmetric system has eigenstates which have definite symmetry; however, symmetry-broken phases are characterized 
%by an exponentially-small splitting between the two lowest eigenstates with opposite parity, thus  allowing symmetry-broken solutions to be approximately-degenerate ground states in the thermodynamic limit. 
For bosonic symmetry-broken systems, there is typically a parameter of the model that controls the separation of  the degenerate ground states or steady states in phase space, which one can tune to reach the ``thermodynamic limit'' \cite{carmichael2015, ciuti2018, lieu2020}. For the case of the photonic cat code, the thermodynamic parameter is the drive strength divided by the two-photon loss strength: $N = \lambda / \kappa_2$; a phase diagram  can be drawn for $\kappa_2=0$ \cite{lieu2020}. For fluxonium, we find that the overlap between between the symmetry-broken states is exponentially small in the parameter:
\begin{align} 
    \langle {-\alpha}, \theta | {+\alpha}, \theta \rangle = \exp[-\frac{1}{2}|2 \alpha'|^2], \\
    \alpha' \equiv \alpha e^{\theta} = \frac{\pi}{2} \left( \frac{E_j}{2 E_c} \right)^{1/4} \left( \frac{E_j}{E_j + E_l} \right).  \label{eq:alpha_p}
\end{align}
Thus $N = |\alpha'|^2$  controls the separation of the wavefunctions in phase space, which is 
%. Note that $N$ is 
mostly determined by the 
%We suggest that for the fluxonium Hamiltonian, the  analogous parameter  is proportional to the 
ratio of the Josephson energy to the capacitive energy, only weakly depending on the inductive energy.
This suggests that the vertical  line characterized by $E_c/E_l=0$ in Fig.~\ref{fig:overlap}(a) is the true phase diagram; by tuning $E_c$ one can tune towards the thermodynamic limit, analogous to adding more lattice sites in the transverse-field Ising model. We can confirm this in Fig.~\ref{fig:overlap}(b) by plotting the lowest four energies of the fluxonium Hamiltonian \eqref{eq:fluxonium_ham} across this boundary: As we approach the thermodynamic limit $E_c\rightarrow 0$, the spectrum is gapped away from the critical point ($E_j = E_l$) and the first excited state is exponentially small in the nontrivial phase. In  Fig.~\ref{fig:splitting} we show that the energy splitting between the ground state and the first excited state  is indeed  exponentially small in the parameter $N$.
%We can indeed confirm that the energy gap is exponentially small in the parameter $N$. (See Fig.~\ref{fig:splitting}.)
In practice,  there will always be some small symmetry-breaking external flux perturbation that will bias the system towards one of the two wells, which will set the frequency ($\epsilon_{01}$) of heavy fluxonium. 
%. The splitting for very heavy fluxonium will be set by the flux inaccuracy that biases the systems towards one of the wells. 
The eigenstates are the symmetry-broken states at small flux offsets.

The squeezed  state solutions  correspond to gaussian wavefunctions in flux space, with a  mean and variance that depends on circuit parameters. 
%are characterized by two real parameters: $\alpha, \theta$. These solutions for the ground state correspond to gaussian wavefunctions in flux space (see Fig.~\ref{fig:fluxonium}, right), which are characterized by their mean and variance. 
The expectation value of the flux operator reads:
\begin{equation}
    \langle  \pm \alpha, \theta |  \phi |  \pm \alpha, \theta \rangle  =     \langle  \pm \alpha | \phi_0 (a+a^\dagger) |  \pm \alpha \rangle  =  \pm 2 \phi_0 \alpha =  \pm \pi \left( \frac{E_j}{E_j + E_l} \right). 
\end{equation}
Thus the squeezed  state solutions have  a flux expectation value that is close to $\pm \pi$, with a deviation that  depends on the ratio of $E_j $ and $E_l$. 
The expectation value of the flux variance (for both squeezed  states) is:
\begin{equation}
    \langle   \phi^2   \rangle  -  \langle   \phi  \rangle^2 =   \phi_0^2 \exp(- 2 \theta) = \sqrt{  \frac{2 E_c}{E_j}   } .
\end{equation}
Thus the flux variance goes to zero as the squeezing parameter is increased, i.e.~the state tends to a delta function  in flux space. 
From these expressions we can infer the expected current through the circuit in the symmetry-broken ground states. The expression for the current through a Josephson junction reads: $I = I_0 \sin \phi$ where $I_0$ is the critical current. For the squeezed  state solutions, we find:
\begin{align}
    I &= I_0 \sin \langle \phi \rangle =  I_0 \sin \left( \pm \pi  \left( 1 -  \frac{ E_l}{E_j + E_l} \right)    \right)  \\
    &\approx \pm I_0 \left( \frac{\pi E_l}{ E_j + E_l} \right) \approx \pm \left( \frac{2\pi}{ \Phi_0}\right) (\pi E_l) = \pm \frac{\Phi_0}{2L}
\end{align}
where we have used $I_0 = E_j(2\pi/\Phi_0), E_l = \Phi_0^2/(4\pi^2 L)$. The squeezed states correspond to clockwise/counterclockwise persistent current states that produce half of a flux quantum through the loop.

\section{Noise bias in fluxonium}

Now that we have found a reasonable parameterization for the ground states of fluxonium as a function of the circuit parameters, we analyze the noise bias in the limit of large $\alpha'$, i.e.~heavy fluxonium. Typical  noise sources in fluxonium couple to the system's charge or flux operators:
\begin{equation}
    \phi = \phi_0 (a + a^\dagger), \qquad     n = n_0(a^\dagger- a), \qquad n_0\equiv \frac{i}{2 \phi_0}.
\end{equation}
In the approximation where we only keep track of the  two (nearly-degenerate) ground states  that span the codespace, we can use Fermi's Golden rule to estimate how the transition rates between states will scale as a function of circuit parameters. The matrix elements of interest obey the following relations:
\begin{align} \label{eq:scaling}
    |\langle - \alpha, \theta |  \phi  | {+ \alpha}, \theta \rangle |^2  &= 0 \\
    | \langle - \alpha, \theta |  n  | {+ \alpha}, \theta \rangle  |^2  &= 4 |n_0 \alpha' e^{\theta}|^2 \exp(-4 (\alpha')^2) \\
     | \langle C_- |  \phi  | C_+\rangle  |^2 &= \pi^2 \left( \frac{ E_j}{E_j + E_l}\right)^2 \label{eq:scaling_c} \\
    | \langle  C_- |  n  |  C_+ \rangle  |^2  &= 4 |n_0 \alpha' e^{\theta}|^2  \exp(-4 (\alpha')^2) \label{eq:scaling_d} 
\end{align}
where we have defined: $\alpha' = \alpha e^{\theta}$ and $| C_\pm\rangle \sim | + \alpha, \theta \rangle \pm | - \alpha, \theta \rangle  $. We again note that the eigenstates  are approximately the  squeezed cat superpositions exactly on the sweet spot, and  the squeezed coherent states just off of the sweet spot \footnote{ If the system is $\delta \phi_e$ away from the sweet spot ($\phi_e = \pi + \delta \phi_e$) then there should be a correction to the expression for $\alpha$ (which sets the average flux value and will shift the left/right wells differently) that will depend linearly on $\delta \phi_e$; however such a correction will not change the Fermi’s golden rule prediction for the scaling of the bit-flip rate or phase-flip rate as a function of the circuit parameters in the heavy limit $E_j \gg E_c$.}.
%(This is generic to symmetry-broken Hamiltonians: In the absence of an explicit symmetry breaking perturbation, the exact eigenstates have definite parity but the gap between the lowest states is exponentially small.)

Let us define the logical 0,1 states to be the states localized in each well, i.e.~$|\pm \alpha, \theta \rangle$, the eigenstates just away from the sweet spot. Then this suggests the following scaling relations for the logical bit and phase-flip error rates (neglecting polynomial factors in front of exponents):
$\text{bit flip}  \sim \exp[-4 (\alpha')^2]$, $ \text{phase flip}  \sim  \text{const.}$. For comparison,  the photonic cat code, obeys: $\text{bit flip}  \sim \exp[-4 |\alpha|^2]$, $ \text{phase flip}  \sim  |\alpha|^2$ where $|\alpha|^2$ is  the number of photons in the cavity. (Note that higher-order corrections change the coefficient in the exponent from 4 to 2 \cite{kirill2024}.) The squeezed photonic cat code \cite{sq_cat1,sq_cat2,sq_cat3} obeys: $\text{bit flip}  \sim \exp[-4 |\alpha'|^2]$, $ \text{phase flip}  \sim  |\alpha|^2$ where $\alpha' = \alpha e^\theta$ and $\theta$ is the squeezing parameter. The squeezed photonic cat code can thus exponentially-suppress bit flips without increasing the phase flips by increasing the squeezing parameter $\theta$ while holding $\alpha$ fixed. In this context, fluxonium naturally realizes a squeezed cat code but has the added benefit that the phase-flip rate does not increase even with $\alpha $,  
%and has the added benefit that  the phase-flip rate does not increase with $\alpha$ 
since it gets canceled with the zero-point flux term: At large $E_j$, $\alpha \sim (E_l / (2 E_c))^{1/4}, \phi_0 \sim (E_l / (2 E_c))^{-1/4} $ such that their product tends to a constant.
%(Note that the photonic squeezed cat code  has the same  error scaling as the conventional cat code with $\alpha\rightarrow \alpha' = \alpha e^\theta$.) This suggests that  the phase flip error rate does not appear to get linearly worse at large $\alpha$, in contrast to the cat code. 
%Thus a repetition code constructed via fluxonium might not suffer from problems at large $\alpha^2$, i.e.~one can make $\alpha$ arbitrarily large without going above the threshold. 
%[We also note that the large $\alpha'$ regime in fluxonium occurs in the ground state of the model in the lab frame, i.e.~it does not cost an extensive energy to reach large $\alpha$ like in the case of the photonic cat code.] 
Note that this analysis depends on a ``two-level system'' description of fluxonium. In what follows we will go beyond this approximation by keeping many fluxonium eigenstates  within the universal Lindblad master equation \cite{nathan2020}, and find some shortcomings in the two-level system analysis in predicting the bit-flip rate.

We perform a dynamical simulation with respect to a concrete noise model to  more accurately quantify the noise bias in fluxonium. We place the external flux in the linear inductor
%briefly change convention by placing the external flux in the linear inductor 
(necessary for a time-dependent flux \cite{you2019, byron2023})
%leading to the Hamiltonian
then shift the variable $\phi \rightarrow \phi - \pi$:
\begin{equation}
    H(t) = 4 E_c(n - \delta n_e(t))^2 +E_l( \phi - \delta \phi_e(t))^2/2 + E_j \cos \phi
\end{equation}
%[Note: this change in convention does not have an effect on the static analysis that we have considered so far; it also does not affect the subsequent Lindblad analysis.] 
with offset charge:  $\delta n_e$ and offset flux: $\pi +  \delta \phi_e $.
\begin{comment}
, leading to the system-bath Hamiltonian:
\begin{align}
H(t) &= H_s + H_{\text{int} } \\
H_s &= 4 E_c n^2 +E_l( \phi - \pi )^2/2 - E_j \cos \phi \label{eq:sys_ham} \\
H_{\text{int} } &= - E_l \delta \phi_e \phi - 8 E_c \delta n_e n
\end{align}
\end{comment}
We model the thermal coupling of fluxonium to its environment via noise operators that follow a Johnson-Nyquist spectral density, which is  typically observed in experiment \cite{aash2010, zhang_2021}:
\begin{align}
    S_{\phi, \phi}(\omega) &= \int_{-\infty}^{\infty} d \tau e^{i \omega \tau } \langle \delta \phi_e(\tau) \delta \phi_e(0) \rangle = c_\phi  \left[ \frac{\hbar \omega}{1- e^{-\beta \hbar \omega}} \right]  \\
    S_{n,n}(\omega) &= \int_{-\infty}^{\infty} d \tau e^{i \omega \tau } \langle \delta n_e(\tau) \delta n_e(0) \rangle = c_n \left[ \frac{\hbar \omega}{1- e^{-\beta \hbar \omega}} \right] 
\end{align}
where $c_{\phi}, c_n$ are constants that depend on the bath.  This ensures that detailed balance is respected, i.e.~$S(\omega) = e^{\beta \hbar \omega } S(-\omega)$. (We do not expect other spectral densities to change the scaling observed below, and comment on the case of $1/f$ flux noise at the end of this section.) 
%(We do not expect other spectral densities to change the scaling observed below. For example, low-frequency flux noise, e.g.~$1/f$, can be approximated by averaging over different flux offset configurations, which will indeed cause dephasing but bit flips will still be exponentially suppressed in the barrier height.)

We proceed to model the dynamics of the system via the universal Lindblad master equation approach, as described in Refs.~\cite{nathan2020, nathan_2024}. This suggests a Lindblad master equation for the dynamics:
\begin{equation}
    \hbar \frac{d \rho}{dt} = \mathcal{L}(\rho) = -i[H, \rho] +  \sum_{j=\phi,n} L_j \rho L_j^\dagger - \frac{1}{2}\{ L_j^\dagger L_j, \rho \}
\end{equation}
with  a single dissipator for each bath:
\begin{align} \label{eq:diss}
    L_\phi &= \sum_{i,j } E_l \sqrt{S_\phi(\omega_{ij})/\hbar  } (\langle j| \phi |i\rangle )
 | j \rangle  \langle i |   \\
 &\equiv  x_\phi \sum_{i,j }  \sqrt{  \frac{\hbar \omega_{ij}}{1- e^{-\beta \hbar \omega_{ij}}}  }  (\langle j| \phi |i\rangle )
 | j \rangle  \langle i | \label{eq:flux_diss} \\
     L_n &= \sum_{i,j} 8 E_c  \sqrt{S_n(\omega_{ij}) /\hbar} (\langle j| n |i\rangle )
 | j \rangle  \langle i | \\
 &\equiv  x_n \sum_{i,j } \sqrt{  \frac{\hbar \omega_{ij}}{1- e^{-\beta \hbar \omega_{ij}}}  }  (\langle j| n |i\rangle ) | j \rangle  \langle i |  \label{eq:charge_diss}
\end{align}
where we define dimensionless constants: $ x_\phi = E_l \sqrt{  c_\phi/\hbar} , x_n = 8E_c \sqrt{  c_\phi/\hbar } $ that characterize the coupling to the bath and thus set the characteristic timescale of the dynamics. In the subsequent analysis, the $x$ coupling constants are held fixed and we focus on how the logical lifetimes scale as a function of circuit parameters (that affect the frequencies and eigenstates) and temperature. The steady state of the model is in principle unique, corresponding to the thermal state of the fluxonium Hamiltonian: $\rho_{ss} \sim \exp[-\beta H]$. In the limit where the system is at or near its sweet spot, this state should have roughly equal support on both of the bottom two wells of the flux potential. However, the mixing time of the dynamics can be exponentially long in the parameter: $E_j / (k_B T)$, i.e.~if we initialize in one well it takes an exponentially-long time to tunnel to the other.

\begin{figure}
    \centering
    \includegraphics[scale=0.2]{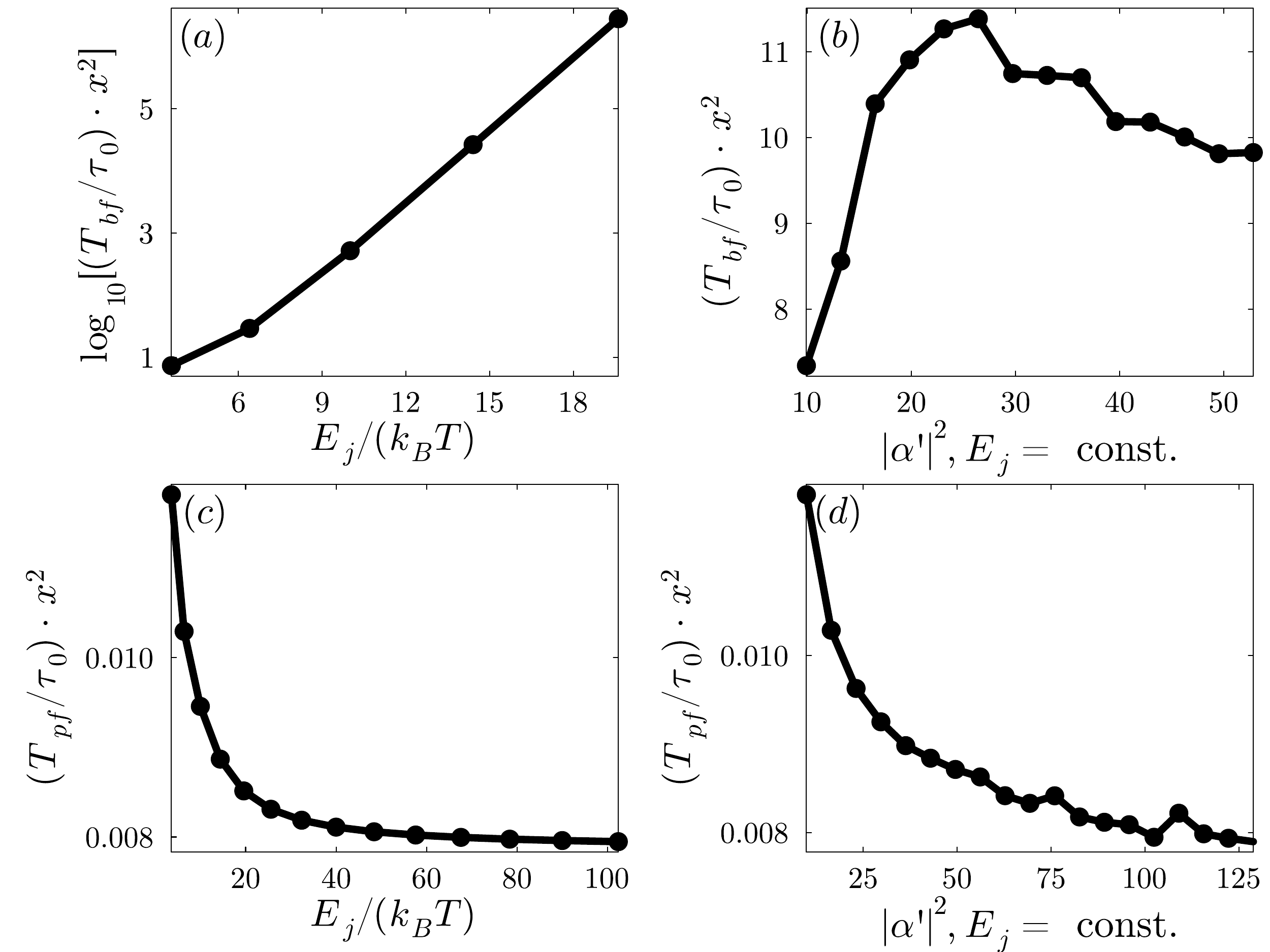} 
    \caption{ Top: Extracted bit-flip time $T_{bf}$ as a function of (a) $E_j / (k_B T)$  and (b) $|\alpha'|^2$ with $E_j$ fixed, multiplied by the dimensionless coupling constant: $ x^2 \equiv x_\phi^2=x_n^2$; times quoted in units of a timescale set by the temperature $\tau_0 = h/ (k_B T)$, approximately a nanosecond for $T=50 \text{ mK}$. Typical values of $x^2$ are on the order of $10^{-5}$ \cite{zhang_2021}. The bit-flip time diverges only in the case of (a).   Bottom:  Extracted phase-flip time $T_{pf}$  as a function of (c) $E_j / (k_B T)$, and (d) $|\alpha'|^2$ with $E_j$ fixed: Both saturate to a constant. Parameters: Flux slightly off sweet spot: $\delta \phi_e/ \pi = 0.03 $; $ k_B T /h =  1 \text{ GHz}$. $E_c/h = 0.1 \text{ GHz}, E_l/h = 0.1 \text{ GHz}$ when varying $E_j$; $E_j/h =   3.6 \text{ GHz}, E_l/h =  0.1 \text{ GHz}$ when varying $E_c$. The dimension of the Hilbert space is varied to include  five eigenstates that have support on both wells.}
    \label{fig:noisebias}
\end{figure}

To obtain a quantitative estimate of the logical bit-flip time, we perform the following simulation: Consider  the fluxonium Hamiltonian in Eq.~\eqref{eq:fluxonium_ham} (just away from the sweet spot $\delta \phi_e /\pi = 0.03 $) with a Hilbert space that is spanned by states with definite flux $|\phi\rangle$ in the range of $\phi \in [-2 \pi, 2 \pi] $. 
%We diagonalize the system Hamiltonian (which is set to be just away from the sweet spot such that the lowest-energy  eigenstates are localized within each well).
The state is initialized in one of the two nearly-degenerate ground states, i.e.~$|\psi_0\rangle \approx  |{-\alpha},\theta \rangle$, localized near the left well at $\phi = -\pi$. It is then evolved with the Lindbladian for a time $t$, and the tunneling probability is computed: $q_\text{tunnel}(t) = \text{Tr}[\rho(t) \Pi_r]$ where $\Pi_r \sim \int_0^{2\pi} d\phi |\phi \rangle \langle \phi |$ is a projector onto $\phi=0$ to $\phi=2 \pi$. This quantifies whether the state has tunneled across the barrier to the other side. (Conceptually one can think of this as doing a readout operation that only distinguishes between states in different wells.) We find that the tunneling (bit-flip) probability obeys the relation: $p_\text{tunnel}(t) = e^{-t/T_{bf}}/2$, i.e.~it decays exponentially with a characteristic timescale $T_{bf}$ towards a state with equal support on both wells. We numerically extract this parameter and observe how it scales with the circuit parameters in Fig.~\ref{fig:noisebias}(a,b).

In Fig.~\ref{fig:noisebias}(a) we find that the bit-flip time $T_{bf}$ indeed diverges as $E_j$ increases for a fixed $k_{B} T$. This agrees with intuition: As the height of the barrier is increased, the time to tunnel across diverges. This is reminiscent of the extensive energy barrier in the classical 2D Ising model, which scales as $\sim M J$ for an $M \times M$ lattice with coupling constant $J$. However we note that fluxonium does not appear to have a finite-temperature phase transition: The mixing time is exponential in $E_j/(k_{B} T)$, as opposed to the Ising model whose mixing time is exponential in linear lattice size $M$ for $T <T_{\text{critical}}$.  This suggests that fluxonium has a well-defined quantum phase transition [evidenced by the spectral analysis in Fig.~\ref{fig:overlap}(b)] but no thermal phase transition, reminiscent of behavior in the 1D transverse-field Ising model or the 2D toric code. Nonetheless, fluxonium is capable of exponentially protecting a classical bit by increasing $E_j$ (akin to increasing the Ising interaction strength in the 1D Ising model) which can be done by decreasing the physical insulating barrier.

In contrast, the bit-flip time $T_{bf}$ saturates if we increase separation between codewords by decreasing $E_c$. [See Fig.~\ref{fig:noisebias}(b).] This represents a shortcoming of the two-level system Fermi's golden rule analysis described in Eq.~\eqref{eq:scaling}: The direct matrix elements between codewords are exponentially suppressed but it is still possible to tunnel from one well to another via intermediate states since the barrier height is not increasing. Similar effects  have been observed in Kerr cats  \cite{noh2022, frattini_2024}. This suggests that more than two levels need to be considered in order to properly account for fluxonium's bit-flip time (at least in certain heavy limits).

To obtain a quantitative estimate of the logical phase-flip time, we perform an analogous simulation: The state is initialized in a positive superposition of the ground states localized in each well, $|\psi_0\rangle \sim  |-\alpha,\theta \rangle+ |\alpha,\theta \rangle$. It is then evolved with the Lindbladian for a time $t$, and the expectation value of the parity operator is computed: $p_{\text{phase}}(t) = \text{Tr}[\rho(t) P]$ where $P$ is the parity operator which takes $P |\phi \rangle =|-\phi \rangle $. We find that the parity decays exponentially: $p(t)=e^{-t/T_{pf}}$ with a characteristic timescale $T_{pf}$. We numerically extract this parameter and observe how it scales with the circuit parameters.

The extracted $T_{pf}$ values are presented in Fig.~\ref{fig:noisebias}(c) for increasing $E_j$, and in Fig.~\ref{fig:noisebias}(d) for  decreasing $E_c$. We find that the $T_{pf}$ time saturates in both cases, in agreement with the prediction from Fermi's golden rule in Eq.~\eqref{eq:scaling_c}. This can be understood intuitively: In the infinitely-squeezed limit ($E_c=0$) where eigenstates are definite flux state $|\phi\rangle$, the dephasing rate is proportional to the energy splitting between the minima of the two wells as a function of the external flux: $d \epsilon_{01}/ d \phi_e = E_l \pi$, which is finite.

\begin{figure}
    \centering
    
    \includegraphics[scale=0.2]{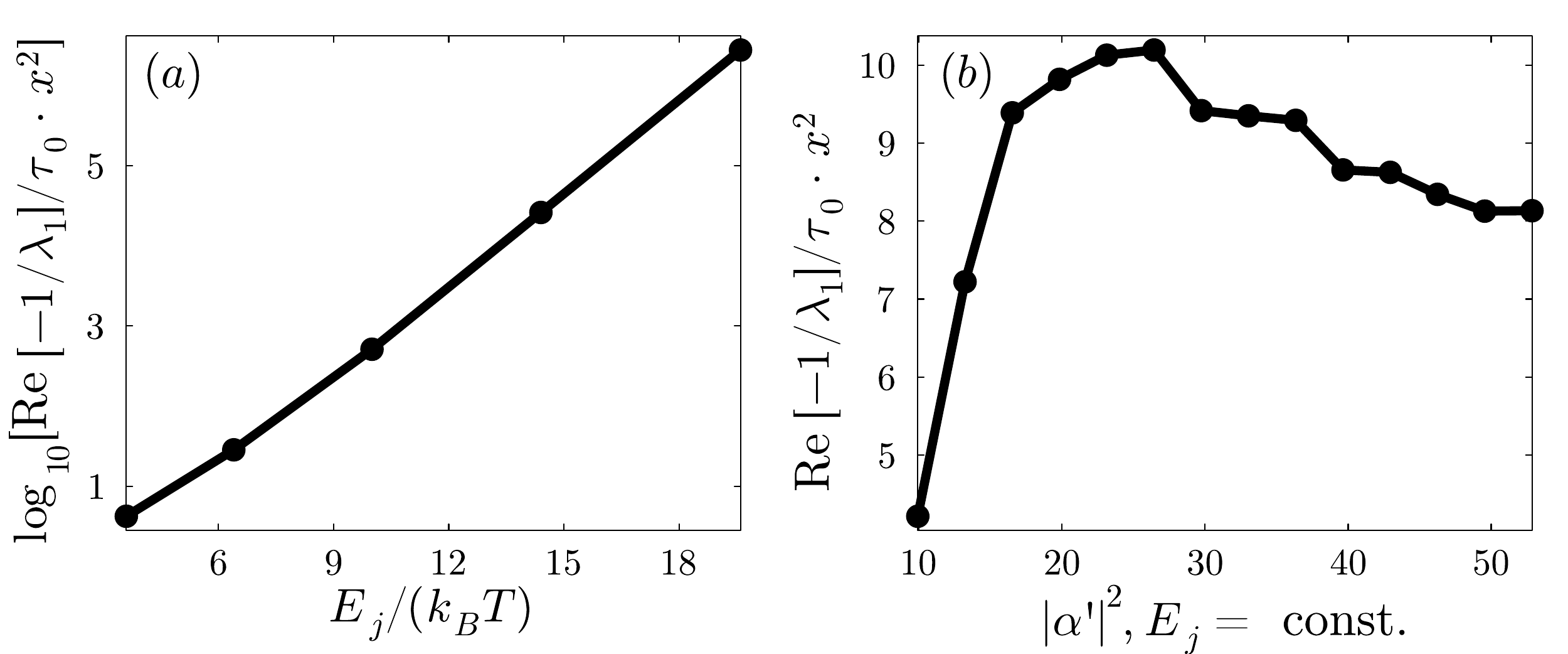} 
    \caption{ Inverse of the real part of the two lowest eigenvalues in the Lindblad spectrum (in units of $\tau_0 = h/ (k_B T)$). The parameters are the same as in Fig.~\ref{fig:noisebias}. Only in  (a) do we find an eigenvalue that is exponentially close to zero, implying a stable classical bit.}
    \label{fig:lind_spec}
\end{figure}

Note that these scaling relations can also be understood by considering the eigenvalues of the Lindbladian. The Lindbladian is guaranteed to have one exact eigenvalue of zero, which we label $\lambda_0=0$, corresponding to the thermal steady state: $\mathcal{L}(\exp[-\beta H] )=0$. We label $\lambda_1$ as the eigenvalue with the next-smallest decay rate. Fig.~\ref{fig:lind_spec} plots $\lambda_1$ for the same parameters as before. Indeed we find the behavior of the $T_{bf}$  is approximately inversely proportional to the corresponding Lindblad eigenvalue. This again suggests a good classical bit in the limit $E_j \gg k_B T$.

Before concluding this section, we discuss a couple of assumptions that could affect the conclusions above. We have assumed that a time-dependent external flux enters the Hamiltonian  via a time-dependent inductive energy and a static Josephson energy, in accordance with the analysis in Refs.~\cite{you2019, byron2023}. If some fraction of the time-dependent flux is instead included in the Josephson energy then this would result in a dissipator whose rate is proportional to $E_j$ (instead of $E_l$), such that the dephasing rate would get linearly worse with the barrier height (resulting in a noise-bias tradeoff similar to the standard cat code). One might need to be careful about how to treat the time-dependent flux if a Josephson junction array is used to approximate the linear inductor.

We have also assumed a Markovian noise model  via a Johnson-Nyquist spectral density. Notably $1/f$ flux noise can be a significant contributor to dephasing in experiments and does not admit a Markovian description. We can use conventional formulas to estimate the decay time for contributions from $1/f$ flux noise. For the dephasing rate \cite{zhang_2021}:
\begin{equation}
    \frac{1}{T_{pf;1/f}} =\sqrt{2} A_{\phi_e} \frac{\partial \epsilon_{01}}{\partial \phi_e} \sqrt{|\ln[\omega_{\text{low}} t_{\text{exp}}] |}
\end{equation}
where  $A_{\phi_e}$ is the amplitude of the noise, $\omega_{\text{low}}$ is an arbitrary low-frequency cutoff, and $t_{\text{exp}}$ is the time of the experiment. In the infinitely-heavy limit ($E_c=0$) one can explicitly solve for the derivative of the spectrum: $\partial \epsilon_{01} / \partial \phi_e = E_l \pi$, which suggests that the phase flip rate will again saturate to a constant (up to the log correction). The bit-flip rate \cite{zhang_2021} is proportional to: $1/T_{bf; 1/f} \sim |\langle \bar{0} | \phi |\bar{1} \rangle |^2/\epsilon_{01}$ which again should be exponentially suppressed due to the matrix element when the system is just off of the sweet spot.

\section{Cat-qubit-inspired gates}

\begin{figure}
    \centering
    \includegraphics[scale=0.75]{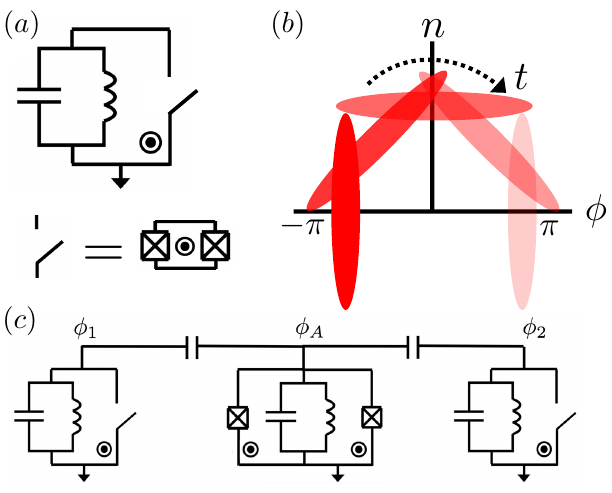} 
    \caption{ (a) A tunable symmetric squid with large $E_{j, \text{max}}$ serves as an effective open/close switch for the circuit. (A closed switch corresponds to $E_j = E_{j, \text{max}}$, i.e.~large critical current.) (b) Sketch of the Wigner function of a squeezed state during the bias-preserving $X$ gate. At $t=0$ the switch is opened and the  $\phi=-\pi$ state (dark red) starts to perform a  rotation in  phase space (lighter red) until it reaches the other codeword ($\phi=\pi$, lightest) at $t=\pi/\omega$. Codeword phase-space separation is maintained throughout the gate. (c) Two tunable fluxoniums are coupled via an asymmetrically-threaded squid (ATS). A $CX$ gate is achieved by flux driving the ATS.}
    \label{fig:flux_gate}
\end{figure}

The  analogy between cat qubits and fluxonium  allows us to translate bias-preserving gate schemes from the former \cite{puri_2020, prx2019, putterman2024}  to the latter.  In this section we briefly sketch a cat-qubit-inspired single-qubit $X$ gate, and two-qubit $CX$ gate  for fluxonium. Consider a tunable $E_j$ fluxonium Hamiltonian (at the sweet spot): 
\begin{equation}
    H(t) =  \hbar \omega a^\dagger a   +  E_j(t) \cos( \phi_0(a+a^\dagger)).
\end{equation}
The tunable $E_j(t)$ can be implemented by replacing the usual junction in fluxonium with a tunable squid (and including an additional flux line) \cite{lin_2018}. We would like an ``ideal switch'' \cite{brooks_2013, siegele_2025, nathan_2024} which could tune the value of $E_j$ between zero and some large value $E_{j,\text{max}}$, i.e.~a symmetric squid. [See Fig.~\ref{fig:flux_gate}(a).] At idling, the junction sits at $E_{j,\text{max}}$. To perform an $X$ gate, we diabatically quench to $E_{j}=0$ such that the Hamiltonian is just the harmonic oscillator: $ H =  \hbar \omega a^\dagger a$.  The squeezed states evolve via: $\exp[-i H t /\hbar ] |\pm \alpha, \theta \rangle = |\pm \alpha e^{-i \omega t  }, \theta e^{-i 2 \omega t}  \rangle$. After waiting a time $t=n \pi / \omega, n \in \text{odd},$ the two codewords evolve into each other, thus performing an $X$ gate while maintaining a large phase-space separation throughout the entire evolution.  [See Fig.~\ref{fig:flux_gate}(b).]
The physical picture is as follows: An ideal switch essentially flips between an open and closed circuit. At idling, the switch is closed, and the persistent current states are stable. To perform a gate, we flip the switch to an open circuit: The current through the inductor starts to charge the capacitor, evolving to a definite charge state at $t=\pi  / (2\omega) $; then the current starts going in the opposite direction through the inductor at $t=\pi  / \omega $ at which point we flip the switch back to  closed.

In practice, the unitary fidelity of this scheme will likely be limited by the minimum Josephson energy $E_{j, \text{min}}$ (determined by how symmetric one can make the junction energies in fabrication) which should obey $E_{j, \text{min}} \ll \hbar \omega$,  and how diabatic one can make the switch (ideally the squid's flux-pulse rise time should obey $t_{\text{rise}} \ll 1/\omega$). An example fluxonium circuit that could meet these conditions has parameters: $E_c/h = E_l/h = 0.5 \text{ GHz} \Rightarrow   \omega = (2\pi )  1.4 \text{ GHz}, E_{j,\text{max}}/h = 10 \text{ GHz} $. A junction mistargetting of $\sim 1\%$ implies $E_{j,\text{min}}/h \sim 0.1 \text{ GHz} $ which obeys $E_{j, \text{min}} < \hbar \omega$. A $50 \text{ ps}$ rise time for the DC flux pulse \cite{ohki_2005} is  compatible with the condition $t_{\text{rise}} < 1/\omega = 100 \text{ ps}$. A numerical simulation (with the parameters described above) suggests that the $X$ gate can be done in less than a nanosecond with a $ 2 \cdot 10^{-4}$ error (probability of not ending up in the opposite well).

Our scheme is similar to a recent proposal for an arbitrary single-qubit gate in fluxonium \cite{siegele_2025}, which uses a similar tunable junction  to quench from a heavy fluxonium regime to a light fluxonium regime (where $E_j \sim \hbar \omega$), instead of the proposed quench to a harmonic oscillator regime (where $E_j \ll \hbar \omega$) described above. While the proposal in Ref.~\cite{siegele_2025} is able to perform an  arbitrary rotation about the $X$ axis (and might be easier to achieve in experiment), it (necessarily)  does not maintain large phase-space separation when acting on squeezed states, e.g.~a $\pi/2$ rotation about the $X$ axis will send squeezed states to cat states. In contrast, the proposal described above can only do an $X$ gate but maintains phase-space separation.  It would be interesting to see the tradeoff between noise bias and arbitrary $X$ rotation fidelity as a function of gate/circuit parameters.

%It would be interesting to investigate which parameter regimes correspond to 
%interpolate between these two regimes (e.g. by lowering $E_{j,\text{min}}$).

%A recent proposal for a similar single-qubit gate scheme using a tunable fluxonium \cite{siegele_2025} considers a quench from a heavy fluxonium regime to a light fluxonium regime (where $E_j \sim \hbar \omega$), instead of the proposed quench to a harmonic oscillator regime (where $E_j \ll \hbar \omega$) described above. The goal in Ref.~\cite{siegele_2025} is to perform arbitrary single-qubit rotations
%to perform an arbitrary $Z$ rotation, 
%instead of the bias-preserving $X$ gate described here.

For a bias-preserving two-qubit $CX$ gate, we consider two tunable fluxoniums (described above) that are capacitively coupled to an asymmetrically-threaded squid (ATS) \cite{lescanne2020} which serves as a coupler. [See Fig.~\ref{fig:flux_gate}(c).] When performing a gate, we set the tunable junction energies of both fluxoniums to zero, causing them to evolve via their harmonic oscillators (analogous to turning off stabilizing dissipation during gates in cat qubits \cite{putterman2024}). At the asymmetric flux bias point, the (dressed) Hamiltonian is \cite{lescanne2020}:
\begin{equation}
    H = \left( \sum_{i=1,2,A} \hbar \omega_i a_{i}^\dagger a_{i} \right) -2 E_{j,A} f(t) \sin( \phi_{1} +\phi_2 +\phi_A  )
\end{equation}
where $1,2$ label the two fluxoniums, $A$ labels the ATS, and $f(t)$ is a flux pump on the ATS. In the rotating frame  of the two fluxoniums, a flux drive on the ATS (at the frequency $\omega_1$) can bring the term $ (a_1 + a_1^\dagger) a_2^\dagger a_2$ on resonance. If the first fluxonium is in a flux state at $t=0$, it is an eigenstate of $\phi_1 \sim a_1 + a_1^\dagger$, such that   the rotating-frame Hamiltonian is: $\epsilon Z_1  a_2^\dagger a_2$ in the logical basis (for some strength $\epsilon$), which will induce a rotation on the target fluxonium (2) conditioned on the state of the control (1)  \cite{sq_cat3, sq_cat2}. Waiting a time $t=\hbar \pi/ (2\epsilon) $ will induce a conditional $X$ gate on the target. We anticipate that the condition $\epsilon \gg E_{j,\text{min}}$ needs to be satisfied to minimize unitary error.

Both the $X$ and the $CX$ gate schemes described above involve turning off the protecting barrier while doing a gate, raising the question of whether errors accumulated during this time would degrade the noise bias.  We  note that recent experiments on cat qubits have demonstrated that exponential noise bias persists even when the stabilizing dissipation (responsible for the protecting barrier) is turned off for a large fraction (e.g.~2/3) of the time. (See Fig.~6 in Ref.~\cite{putterman2024b}.) This suggests that gate schemes that turn off the potential barrier during a gate can still be compatible with exponential noise bias, provided that phase space separation between codewords is maintained throughout the gate.

\section{Noise bias in the  $\cos(2 \theta)$ qubit}

We expect the same qualitative behavior to occur in other protected qubits that have a double-well flux potential that leads to nearly-degenerate ground states due to $\mathbb{Z}_2$ spontaneous symmetry breaking. Indeed the Lindblad simulations demonstrating noise bias above were done in the flux basis of a double-well potential, which should closely resemble the dynamics of other protected qubits such as the  $\cos(2 \theta)$ qubit or the zero-pi qubit. (Note that the zero-pi qubit can be viewed as a circuit with an effective $\cos(2 \theta)$ potential \cite{Paolo_2019}.) The $\cos(2 \theta)$ qubit has a  Hamiltonian:
\begin{equation}
    H_{\cos(2 \theta)} = -E_{j2} \cos(2 \theta)-E_{j1} \cos( \theta - \phi_e)  + 4 E_c (n-n_e)^2 
\end{equation}
with $E_{j2} \gg E_{j1}$.  Physically this corresponds to a Josephson junction that primarily allows \textit{pairs} of Cooper pairs to tunnel across the barrier, at a rate $E_{j2}$, which can be realized in a number of ways, including with  semiconductors \cite{larsen2020},  d-wave superconductors \cite{patel2024, vool2024}, or with conventional circuit elements \cite{cos2t, kitaev:2006zeropi, brooks_2013, pop_2008, bell_2014}.  The precise way that charge and flux noise enters the Hamiltonian depends on the exact implementation of the model, however we have specialized to the case of a rhombus of Josephson junctions with half of a flux quantum through the loop, with $\theta$ representing the diagonal phase difference \cite{pop_2008, bell_2014}.
%a d-wave superconductor on top of an s-wave superconductor \cite{patel2024}. 
[Placing the time-dependent flux in the $\cos(\theta)$ term is the worst case for both bit flips and phase flips, i.e.~it induces an energy splitting between the two minima.]

One difference with respect to fluxonium is that the $\cos(2 \theta)$ qubit has a compact phase variable, meaning that $\theta$ is only well defined between the interval of $[0, 2\pi]$, corresponding to the Hilbert space of a rotor \cite{Albert_2017,vuillot_2024}. This implies that charge states are labeled by discrete integers while phase states  assume a continuous value, related via:
\begin{equation}
    |\theta\rangle = \frac{1}{\sqrt{2\pi}} \sum_{n\in \mathbb{Z}} e^{-i \theta n}|n\rangle, \qquad |n\rangle = \frac{1}{\sqrt{2\pi}} \int_0^{2\pi} e^{i \theta n}|\theta \rangle d\theta.
\end{equation}
In the heavy limit, the nearly-degenerate ground states have definite phase (indefinite charge) centered at the bottom of each well: $|\bar{0} \rangle \approx| \theta=0 \rangle $, $|\bar{1} \rangle \approx| \theta=\pi \rangle $ with an overlap that is exponentially small in $E_{j2} / E_c$.  The Hamiltonian (at the sweet spot $E_{j1} = 0$) has a Cooper-pair parity symmetry ($[H, \exp[i \pi n]] = 0$) but the logical codewords spontaneously break this symmetry: 
\begin{align}
    \exp[i \pi n] |\bar{0} \rangle &\sim  \exp[i \pi n]  \left[\sum_{n \in \text{even} }|n \rangle + \sum_{n \in \text{odd}} |n \rangle\right] \\
    &\sim \sum_{n \in \text{even}} |n \rangle - \sum_{n \in \text{odd} } |n \rangle \sim |\bar{1} \rangle .
\end{align}
The $X$ eigenstates $|\pm\rangle \sim |\bar{0} \rangle \pm |\bar{1} \rangle $ represent states with a uniform superposition of even and odd charge imbalance respectively. They transform into each other in the presence of single Cooper pair tunneling: $\cos(\theta) |+\rangle \sim |-\rangle$. Thus the logical phase-flip error rate is directly proportional to the rate of incoherent single Cooper pair tunneling (similar to single-photon-loss-inducing phase flips in the  cat code). 

\begin{figure}
    \centering
    \includegraphics[scale=0.2]{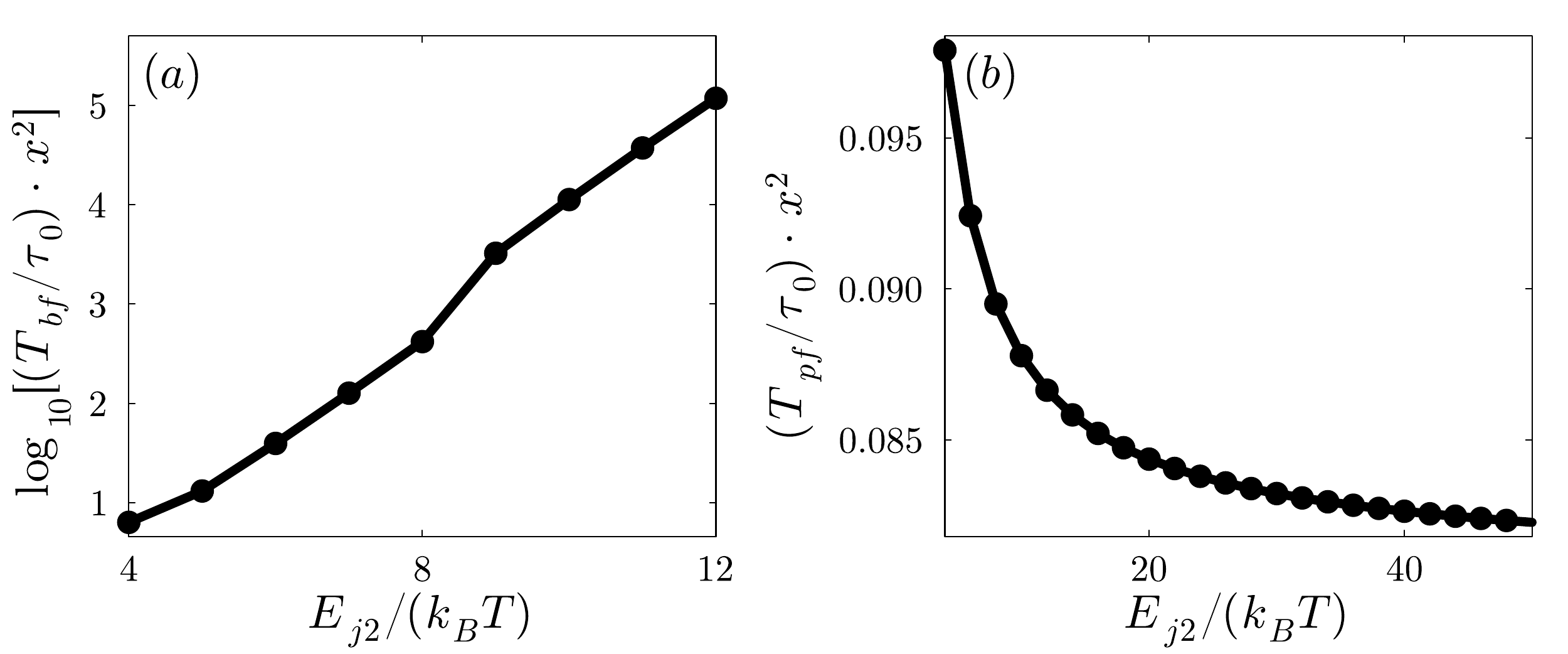} 
    \caption{  Extracted  (a) bit-flip and (b) phase-flip times for the $\cos(2 \theta)$ qubit multiplied by the dimensionless coupling constant: $x^2 \equiv x_{\cos(\theta)} ^2= x_n^2 $. Dissipators are of the form in Eqs.~\eqref{eq:flux_diss} and \eqref{eq:charge_diss}, with the system operators $\cos(\theta)$ (found by Taylor expanding for small deviations in the external flux \cite{le_2019}) and $ n$ respectively. Parameters: Times quoted in $\tau_0 = h/ (k_B T)$. $ k_B T /h =  1 \text{ GHz}$, $E_c/h = 0.1 \text{ GHz}$. }
    \label{fig:cos2t}
\end{figure}

%These correspond to states with definite phase: 
In analogy with fluxonium, we expect exponentially-good bit-flip protection in the parameter $E_{j2}/ (k_B T) $. The logical phase-flip rate should saturate to a constant due to flux-noise induced dephasing which causes a time-dependent energy splitting  between the two lowest-energy states. We can indeed confirm this via an analogous Lindblad master equation simulation, again assuming a Johnson-Nyquist spectral density for charge and flux noise. The results are provided in Fig.~\ref{fig:cos2t} and indeed confirm the aforementioned scaling.  [We  note that an approach to generate a protected $X$ gate in a $\cos(2 \theta)$ qubit was discussed in Ref.~\cite{leroux2023} which involves adding an additional circuit degree of freedom (node).]

%The main source of noise  comes from an explicit symmetry-breaking term in the Hamiltonian $E_{j1}$ which allows single Cooper pairs to tunnel across the barrier. If an environmental degree of freedom   couples to the $E_{j1}$ term then this causes a time-dependent energy splitting to occur between the two lowest-energy states. This dephases the qubit, in analogy   to the flux-noise-induced dephasing in fluxonium described above.

\section{Towards a 2D  Ising model with protected qubits}

\begin{figure}
    \centering
    \includegraphics[scale=0.5]{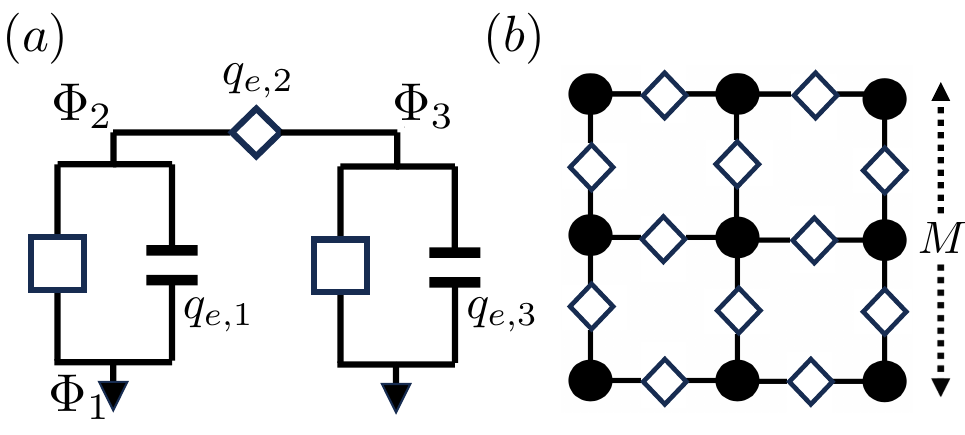} 
    \caption{ (a) Circuit diagram of two  $\cos(2 \theta) $ qubits coupled via a quantum phase slip junction (diamond).  (b) A 2D lattice of $\cos(2 \theta) $ qubits   (black dots) are connected via quantum phase slip junctions (diamonds) that generate an $XX$ interaction in the logical basis. This interaction causes phase-flip errors to be exponentially suppressed as a function of the linear lattice size $M$ at low temperature $T<T_c$.}
    \label{fig:qps}
\end{figure}

The noise bias in protected qubits provides a way to exponentially suppress the logical bit-flip error rate without the need of active error correction, e.g.~by increasing $E_j/(k_B T)$ (decreasing the physical insulating barrier).  This opens up the possibility of  architecture choices that exploit the noise bias, including: (A) a 2D rectangular surface code \cite{brown_2021, amazon_cat,hann_2024}, (B) a 1D repetition code \cite{prx2019, putterman2024}, (C) a 2D fully passive quantum memory \cite{lieu2023candidate}. In (B) and (C) we leverage the idea that the logical bit-flip rate can be made arbitrarily small at the level of a single fluxonium, then we use the ``outer code'' only to correct the  logical phase-flip error.  In (C), the  protection of the repetition code is done without measurements via an Ising interaction on a 2D lattice, e.g.~a ferromagnet \cite{eric2002, liu2023dissipative}: An interaction of the form: $-J \sum_{\langle i j \rangle}X_i X_j$ between nearest neighbors will ensure that logical phase flips are exponentially suppressed in linear lattice size provided $k_{B}T < 2.27 J$. (An $XX$ interaction is needed to suppress dephasing in this basis convention.) To realize (C) we would thus  like to create an Ising interaction between protected superconducting qubits on a 2D lattice, similar to an analogous proposal for cat qubits in SM Sec.~5 of Ref.~\cite{lieu2023candidate}. [See Fig.~\ref{fig:qps}(b).] The logical error rates should then obey the following scaling relations
\begin{align}
    \text{logical bit flip rate } &\sim M^2 e^{-c E_j/(k_B T)}, \\
    \text{logical phase flip rate } &\sim  e^{-c' M},
\end{align}
where $M$ is the linear size of the lattice and $c,c'$ are constants. Note that the bit-flip rate increases quadratically in $M$ since a bit flip on any site will cause a logical bit flip.  As $ E_j ,M \rightarrow \infty$  both logical error rates are exponentially suppressed.  

Ref.~\cite{epstein2022} has shown that it is possible to generate an $XX$ interaction between neighboring fluxonium qubits by using a bifluxon element as a coupler, with strengths of order $5 \text{ GHz}$ as found via numerical simulations. (A fridge at $50 \text{mK}$ corresponds to an energy scale of $1 \text{ GHz}$, well below the critical temperature.) An interesting question remains to characterize the effective $XX$ interaction strength as a function of generic circuit parameters, particularly  in a 2D architecture.

Here we would like to consider a simpler circuit design  to generate an  $XX$ interaction between protected $\cos(2 \theta)$ qubits  via quantum phase slip junctions \cite{mooij_2006, manu2012, serniak_2024, ardati_2024, purmessur2025}.  We consider the circuit depicted in Fig.~\ref{fig:qps}(a), i.e.~a pair of $\cos(2 \theta)$ qubits which are connected via a quantum phase slip junction between the upper nodes. In the presence of nonlinear charge and flux elements, conventional methods of circuit quantization \cite{vool2017, Ciani_2024, rasmussen_2021}  cannot be applied and we instead rely on the newly-developed description of symplectic quantization to arrive at the quantized Hamiltonian of the system \cite{osborne2024, osborne2024b, divincenzo2023, parra_2024}, closely following the procedure  in Ref.~\cite{osborne2024}. In this approach, charge degrees of freedom are associated to each edge of the graph, while flux degrees of freedom are associated with each node. The capacitive and inductive energies can then be expressed as:
\begin{align}
    \bar{E}_c &= \frac{1}{2C} (q_{e,1}^2 + q_{e,3}^2 )-E_q \cos( \kappa_q q_{e,2}) \\
    \bar{E}_i &= -E_j [ \cos( 2 \kappa_\phi (\Phi_2 - \Phi_1)  ) + \cos(2 \kappa_\phi (\Phi_3 - \Phi_1))]
\end{align}
where we have defined $\kappa_q = 2\pi/(2 e),\kappa_\phi = 2\pi / \Phi_0 $, and assume ideal $\cos(2 \theta)$ qubits. The Lagrangian of the system can then be expressed as:
\begin{align}
    \bar{L} &= q_{e,1}( \dot{\Phi}_2 - \dot{\Phi}_1) + q_{e,2}( \dot{\Phi}_3 - \dot{\Phi}_2)+ q_{e,3}( \dot{\Phi}_1 - \dot{\Phi}_3) - \bar{E}_c - \bar{E}_i \nonumber \\
    &= (q_{e,1} - q_{e,2})( \dot{\Phi}_2 - \dot{\Phi}_1) + (q_{e,3} - q_{e,2})( \dot{\Phi}_1 - \dot{\Phi}_3)- \bar{E}_c - \bar{E}_i \nonumber \\
    &\equiv Q_{e,1} \dot{\Phi}_{e,1}+Q_{e,3}\dot{\Phi}_{e,3}- \bar{E}_c - \bar{E}_i .
\end{align}

The Hamiltonian should have only two real degrees of freedom after removing redundancies from Kirschoff's laws, which we take to be $Q_{e,1},Q_{e,3}$ (and conjugate variables). We use the following constraint \cite{osborne2024} to reduce the number of degrees of freedom in $\bar{E}_c, \bar{E}_i$:
\begin{equation}
\sum_j \frac{\partial \bar{E}_c}{ \partial q_{e,j}} = 0 \implies \frac{q_{e,1} +q_{e,3}  }{C} + E_q \kappa_q \sin(\kappa_q q_{e,2}) = 0 .
\end{equation}
Rewriting these equations in terms of the real degrees of freedom leads to a nonlinear equation:
\begin{equation}
    -\frac{(Q_{e,1} + Q_{e,3)}}{4e} = \frac{q_{e,2}}{2e} + \left( \frac{\pi E_q}{8 E_c} \right) \sin( \kappa_q q_{e,2})  \label{eq:kepler}
\end{equation}
which is called Kepler's equation \cite{divincenzo2023}. In general it does not have an analytic solution for $q_{e,2}$ in terms of $Q_{e,1},Q_{e,3}$. A nonlinear constraint is associated with a ``singular'' circuit, which can be remedied by placing a linear inductor in series with the quantum phase slip junction (representing a parasitic inductance). We will analyze this remedy below, but for now we can  make further progress by  assuming that $E_q \ll E_c$ such that we can drop the sine term. This suggests:
\begin{equation}
   q_{e,2} \approx    -\frac{(Q_{e,1} + Q_{e,3})}{2} .
\end{equation}

With this expression, the Hamiltonian of the model reads:
\begin{align}
    H = &\frac{1}{2C} \left( \frac{Q_{e,1}^2 + Q_{e,3}^2}{2}  - 2 Q_{e,1} Q_{e,3} \right) \\
    & -E_q \cos(\kappa_q (Q_{e,1} + Q_{e,3})/2) \label{eq:nl} \\
    & -E_j[\cos(2 \kappa_\phi \Phi_{e,1} ) + \cos(2 \kappa_\phi \Phi_{e,3} )  ].
\end{align}
This describes the  $\cos(2 \theta)$ Hamiltonian of two qubits, along with a standard charge-charge capacitive interaction, and a nonlinear capacitive interaction. Note that the nonlinear capacitive interaction \eqref{eq:nl} can be expressed as a sum of exponentials, which act on states as: 
\begin{equation}
    \exp[\pm i 2 \pi ( n_1 + n_3 )/2 ]| \phi_1, \phi_2 \rangle = | \phi_1 \pm \pi, \phi_2 \pm \pi \rangle
\end{equation}
where we have defined the dimensionless charge and flux operators via: $n_{e,j} = Q_{e,j}/(2e), \phi_{e,j} = \Phi_{e,j} / [\Phi_0/(2\pi)]$.
The logical codewords are defined to be eigenstates localized at the bottom of the two wells of the flux potential: $|\bar{0} \rangle =| \phi =0 \rangle, |\bar{1} \rangle =| \phi =\pi \rangle $. Since the Hilbert space is compact (the phase is only well defined mod $2\pi$),   this implies that the nonlinear interaction causes the desired $XX$ interaction in the logical basis.  

To cure the nonlinear constraint \eqref{eq:kepler} (needed to consider the regime $E_q > E_c$) we can split the node $\Phi_3$ into two nodes: $\Phi_3$ on the left  and $ \Phi_4$ on the right,   separated by a linear inductor $L$. This produces another real degree of freedom in the circuit. Repeating the circuit quantization procedure \cite{osborne2024}, we arrive at the Hamiltonian:
\begin{align}
    &H = \frac{1}{2C}(Q_{e1}^2 + Q_{e3}^2) - E_q \cos(\kappa_q Q_{e2})  \\
    & - E_j[\cos(2 \kappa_\phi \Phi_{e1} ) +\cos(2 \kappa_\phi \Phi_{e3} ) ]+\frac{1}{2L}(\Phi_{e1} +\Phi_{e2}+\Phi_{e3} )^2 .\nonumber
\end{align}
We can gain intuition on this model by examining the limit $C\rightarrow \infty$ (heavy $\cos(2 \theta)$ qubits) and $L\rightarrow 0$ (small parasitic inductance). In the limit where the inductive energy  $\Phi_0^2/L$ is the largest energy scale in the Hamiltonian,  we assume that the relevant states will have flux variables that obey the constraint: $\Phi_{e1} +\Phi_{e2}+\Phi_{e3} = 0$ to minimize the inductive energy, i.e.~fast thermal relaxation to minimize this energy term. The limit $C\rightarrow \infty$ suggests that there is no kinetic energy hence eigenstates have definite flux.  Moreover the Josephson terms ensure that $\Phi_{e1}$ and $\Phi_{e3}$ should sit at the minimum of the cosine potentials.  The relevant states  that span this low-energy manifold can be parameterized via integers $n,m$:
\begin{equation}
    |n,m\rangle \equiv | \Phi_{e1} = n \Phi_0/2, \Phi_{e3} = m \Phi_0/2; \Phi_{e2} = -(n+m)\Phi_0/2  \rangle \nonumber
\end{equation}
We can define logical states:
\begin{align}
    |\overline{0,0}\rangle &\sim \sum_{n,m \in \text{even}} |n,m\rangle, \qquad |\overline{1,1}\rangle \sim \sum_{n,m \in \text{odd}} |n,m\rangle \nonumber \\
|\overline{0,1}\rangle &\sim \sum_{n \in \text{even},m \in \text{odd}} |n,m\rangle, \qquad |\overline{1,0}\rangle \sim \sum_{n \in \text{odd},m \in \text{even}} |n,m\rangle \nonumber .
\end{align}
The nonlinear capacitive term will induce hopping by $\Phi_0$ in the variable $\Phi_{e2}$, which  will cause transitions between  $|\overline{0,0}\rangle \leftrightarrow |\overline{1,1}\rangle$ and $|\overline{0,1}\rangle \leftrightarrow |\overline{1,0}\rangle$, i.e.~an effective $XX$ interaction in this low-energy basis. 

%In the future it would be interesting to consider other coupling schemes to get rid of the standard charge-charge coupling term, and remove the restrictions necessary for an analytic solution to Kepler's equation.
%include a $\sigma_i^+ \sigma_j^- +\sigma_i^- \sigma_j^+ $  interaction between $\cos(2 \theta)$ qubits for the Ising interaction. (It would also be useful to remove the restrictions necessary for an analytic solution to Kepler's equation.)

\begin{comment}
\begin{figure}
    \centering
    \includegraphics[scale=0.4]{classical_bits.pdf} 
    \caption{ Different ways to obtain a good classical bit without active error correction: bosonic qubits,  protected superconducting qubits, and self-correcting lattice models can host stable bits. The listed examples  can all be understood in the language of $\mathbb{Z}_2$ spontaneous symmetry breaking in a Lindbladian.}
    \label{fig:classical}
\end{figure}
\end{comment}

\section{Outlook}

We have studied the protected qubit fluxonium in its Fock basis in order to draw parallels with the bosonic cat qubit. We found analytical expressions for the symmetry-broken ground states of heavy fluxonium, and a protected classical bit due to $\mathbb{Z}_2$ symmetry breaking of the flux potential corresponding to persistent current states in the clockwise/counterclockwise direction. The bit-flip time grows exponentially  as a function of $E_j/ (k_{B} T)$, while the phase-flip time saturates to a constant  with this ratio. Our analysis provides another  example of $\mathbb{Z}_2$ spontaneous symmetry breaking in a Lindbladian leading to a good classical bit. Other examples can be found in bosonic qubits \cite{lieu2020} and self-correcting lattice models \cite{liu2023dissipative}. 
%[See Fig.~\ref{fig:classical}.]
Note that the associated Lindbladians  can either be thermal (e.g.~in the 2D Ising model and fluxonium) or nonthermal (e.g.~the cat qubit). Thermal Lindbladians have the advantage that they arise when coupling a  Hamiltonian to a generic thermal bath (i.e.~they do not require dissipative engineering).

%We note that even with active error correction, one requires a lattice of bosonic modes \cite{noh_2022} to achieve exponential suppression of both logical bit flips and phase flips. It therefore seems unlikely that passive schemes can exponentially suppress both noise flavors in a single mode (or even a few). However our analysis points to several examples of $\mathbb{Z}_2$ symmetry breaking in a single bosonic mode that leads to   a classical bit without measurements. One can then aim to engineer an Ising interaction between such qubits to correct the other error. 

%A generalization of this idea can also be used to do a protected CX gate. For fluxonium, obtaining an imaginary $\alpha$ value would correspond to states with localized charge but not flux, which could 
%For fluxonium, an analogous process could be done with a Hamiltonian that has a tunable JJ to QPS ratio in its Hamiltonian. Future work should explore different schemes to do bias-preserving logical operations with fluxonium. [We note that there is already some work in this direction [].]

Future work should continue to analyze and simplify schemes aimed at generating an Ising interaction between protected superconducting qubits in order to obtain a passive quantum memory. For example, it would be useful to construct a similar model that does not rely on an ideal quantum phase slip element (nonlinear capacitor) since  such elements do not currently have a standardized fabrication process  (in contrast to nonlinear inductors).  At a high-level, the aim is to combine ideas from protected qubits and self-correcting lattice models to engineer a \textit{static two-body interaction}  between qubits on a \textit{ 2D lattice}  such that the energy landscape of the resulting Hamiltonian leads to passive suppression of bit flips and phase flips when coupled to a generic thermal bath. 
%that exponentially suppresses  both logical bit flips and phase flips (as a function of a qubit's phase space separation and the number of lattice sites). 
While some prior studies have asked similar questions \cite{ioffe_2003,ioffe_2012,ioffe:2002}, they have focused on ``zero-temperature'' quantum stability while we would like to identify thermal stability. We note that even with the power of active error correction, one requires a 2D lattice of bosonic modes \cite{noh_2022} to achieve exponential suppression of both logical bit flips and phase flips. It therefore seems unlikely that passive schemes can exponentially suppress both noise flavors to arbitrary precision in a single bosonic mode (or a constant number of modes) since this would imply that passive schemes are more powerful than active ones. 

Another natural question is how to do gates within a fully passive error-correcting device. The gate proposals  should face a similar challenge as those for the 1D repetition cat code \cite{ prx2019, putterman2024}: Gates need to preserve the exponential bit-flip bias at all times. The main advantage of the fully passive scheme (compared to the 1D active repetition code) is that it does not require gates for syndrome rounds,  hence some of the typical speed requirements that are needed  to reach the active error-correcting threshold can be relaxed; only logical gates need to be implemented.
%If one imagines that an Ising interaction between protected qubits can be achieved, the  next question is how to do gates necessary for quantum computation. 
%Another interesting question is how to do gates within a fully passive memory. 
%The advantage of passive error-correcting schemes is that they do not require gates for syndrome rounds,  hence some of the typical speed requirements that are needed  to reach the threshold can be  relaxed. The disadvantage is that all  gates need to preserve the exponential bit-flip bias at all times. 
One important difference between fluxonium and the cat qubit is that the latter has a  set of bias-preserving gates needed to do logical protected computation \cite{puri2017, albert_2016, prx2019, gautier_2023}. (This has been emphasized in Refs.~\cite{schuster2021, epstein2022}.) For example, to do a protected $X$ gate with cat qubits, it suffices to adiabatically change the phase of the drive by $\pi$, thus moving the state $|\alpha e^{i \theta(t) } \rangle$ from $+\alpha$ to $-\alpha$ by tuning the drive phase $\theta$ from $0$ to $\pi$.  In this paper we have sketched a few approaches for performing analogous bias-preserving gates with fluxonium, but a more thorough investigation is needed to assess their experimental potential.
%An open question remains how to do protected logical operations with fluxonium in a similar manner, since it is not straightforward  to do a complex $\alpha$ rotation in phase space via a time-dependent external charge or flux drive. 
Considering the bosonic analogy described in this work might  lead to a  set of bias-preserving gates that can be used for logical protected computation in protected superconducting qubits.
%lead to new  schemes for bias-preserving logical operations with protected qubits.
%Another interesting direction involves  asking how to generate an Ising interaction between protected superconducting qubits in order to obtain a fully passive quantum memory. At a high-level, the question is whether it is possible for a two-dimensional lattice of interacting protected superconducting qubits to exponentially suppress  both logical bit flips and phase flips as a function of a qubit's phase space separation, and the number of lattice sites in the system. While some prior studies have asked similar questions \cite{ioffe_2003,ioffe_2012,ioffe:2002}, they have focused on ``zero-temperature'' quantum stability while we would like to identify thermal stability.

\begin{acknowledgments}
We thank David Schuster, Andrew Osborne,  Yoni Schattner,  Bright Ye, and Christian Siegele for inspiring discussions. We thank Simone Severini, Peter DeSantis, Oskar Painter, Fernando Brand\~{a}o, Eric Chisholm and AWS for supporting the quantum computing program.
\end{acknowledgments}

\bibliography{refs}
\bibliographystyle{apsrev4-1}

\end{document}